# Compressive hyperspectral phasor imaging with single-pixel detection for spectral tasks

Jiaqi Song[1,4], Baolei Liu [2,3,*], Muchen Zhu[1], Yao Wang[1], Yue Yu[2], Zhaohua Yang[2,3], Xiaolan Zhong[1], Fan Wang[1,*]

[1] School of Physics, Beihang University, Beijing 100191, China
[2] School of Instrumentation Science and Optoelectronic Engineering, Beihang University, Beijing 100191, China
[3] Hangzhou International Innovation Institute, Beihang University, Hangzhou, 311115, China
[4] Institute of Physics, China Academy of Sciences, Beijing 100190, China

These authors contributed equally to this work: Jiaqi Song and Baolei Liu
* Correspondence: liubaolei@buaa.edu.cn, fanwang@buaa.edu.cn

**Abstract**

  Spectral vision task plays a pivotal role in extracting discriminative spectral-spatial features from high-dimensional data, enabling fine-grained identification beyond human vision. However, traditional methods usually involve first collecting rich spectral-spatial information and then using complex algorithms to digitally process it into scene classification and recognition, which poses challenges for in data acquisition and processing. Here, we demonstrate a compressive Hyperspectral Phasor Imaging with Single-pixel detection (HyPIS) that leverages highly compressed spatial-spectral data to achieve spectral task. Two optical encoders are used for wavelength-dependent sine- and cosine-encoding that transforms spectral signals into a two-dimensional (2D) phasor plot. By applying spatial-temporal illumination patterns, a single-pixel detector is enough to reconstruct the phasor image of the object. This allows to directly generate pixel-wise spectral task, bypassing 3D hyperspectral data. Our experiments show that HyPIS can perform real-time classification and recognition tasks of different scenes, reducing the required amount of data by two orders of magnitude, and it can still accurately classify under low light and uneven lighting conditions. This work develops a completely new spectral technology that enables spectral tasks to be performed without obtaining high-resolution hyperspectral datasets, holding promise for spectral applications in mobile devices, robotics, and satellite technologies.

**Introduction**

  Spectral vision combines traditional imaging and spectral detection, enabling the acquisition of object information in both spatial and spectral dimensions[1-5]. This capability enables the capture of subtle spectral discrepancies. Therefore, it has been widely applied in many fields, such as scientific research, non-destructive food testing, biomedical applications, industrial inspection, and consumer electronics[6-10]. After the acquisition of high-dimensional spatial-spectral data, versatile machine vision tasks, such as scene classification, identification, and quantitative inversion, are applied for practical applications[6, 11-15]. However, the acquisition of spatial images with hundreds of continuous spectral information leads to explosive data volumes, challenging both data transmission, storage, and processing.

To mitigate the high dimensionality of hyperspectral data, typical approaches such as linear unmixing, principal component analysis, and nonlinear manifold learning methods have been proposed for dimensionality reduction[14, 16-18], aiming to remove redundant spectral information and improve computational efficiency. More recently, a transformative shift toward "spectral machine vision" has emerged, aiming to bypass the heavy data bottleneck entirely by integrating analysis directly into the hardware. A landmark example is the development of Spectral Kernel Machines (SKMs)[19], which use electrically tunable photodetectors to perform classification and recognition of object at the data acquisition points, demonstrating absolute advantages in speed and energy efficiency. However, this reliance on prior knowledge means that the system's performance is essentially related to the scope of its training library, which may pose limitations when encountering completely unknown or rapidly changing spectral environments.

In search of a more universal and training-free alternative, spectral phasor analysis offers an elegant physical-geometry approach[20-22]. By transforming a continuous spectrum into a single point in a 2D phasor plane through Fourier transform, this method allows for rapid spectral unmixing even with low signal-to-noise inputs without using the entire spectrum[23]. To date, this approach has been demonstrated across multiple imaging modalities, such as fast multi-color microscopy, bioluminescence fluorescence microscopy, and snapshot hyperspectral microscopy[24-28]. However, like SKMs, current methods can only reduce the measurement number along the spectral dimension, leaving the spatial data bottleneck unaddressed. Although point scanning can be used to probe different spatial locations[24], this method is inefficient and not conducive to real-time operation.

To bring this gap, we introduce Single-pixel imaging (SPI), an emerging computational imaging technique, to provide a crucial element of spatial compression. It can reconstruct object images under compressed sampling conditions [29-34], offering an effective solution for imaging in unusual wavebands where array detectors are difficult to manufacture or cost-prohibitive[35-37]. To date, it has been widely applied in a wide waveband that ranges from X-rays to terahertz[38-44]. While hyperspectral SPI has been explored, it usually requires complicated spectral modulation and significant acquisition time, with reconstruction processes that remain computationally heavy for video-rate imaging[10, 45, 46].

In this work, we propose a compressive Hyperspectral Phasor Imaging with Single-pixel detection (HyPIS) for spectral vision task which synergizes optical phasor encoding with compressive SPI. Unlike previous spectral task systems, our approach achieves a 2D compression: spectral information is compressed into the coordinates of the 2D phasor plane through optical sine/cosine encoding, while spatial information is compressed modulated by spatial-temporal illumination patterns. The phasor images can be reconstructed by compressive Phasor-SPI algorithm (see *"Principles of HyPIS"*) and perform spectral tasks. With the data compression in both spatial and spectral dimensions, a reduction of the data volume by two orders of magnitude can be achieved, which decreases the requirement for both data transferring and storage capacity. We validate the method in both theoretically and experimentally, with applications for real-time classification and recognition tasks of different scenes. Essentially, HyPIS achieves similar spectral task perception through purely optical encoding, but it eliminates the need for any training dataset while enabling real-time pixel-by-pixel spectral tasks. This work provides a simplified hyperspectral phasor imaging method without acquiring the tremendous 3D hyperspectral image data and the use of complex processing algorithms for spectral tasks. Moreover, our method can be easily adapted for other wavebands, such as near-infrared, mid-infrared, and terahertz.

**Principle**

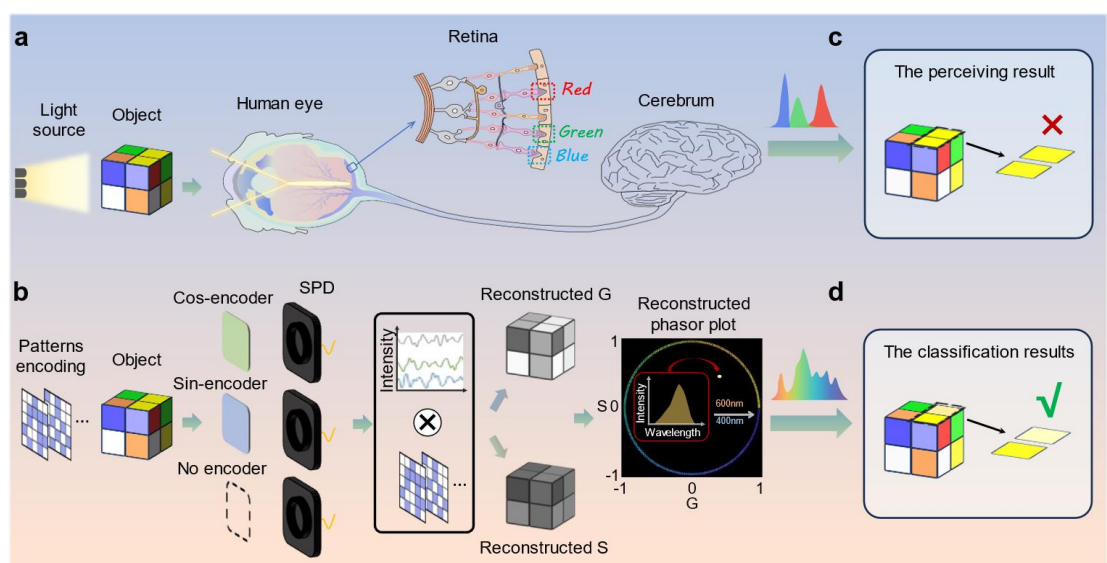

**Fig. 1. Schematic illustrating how HyPIS mimics human vision to enable hyperspectral image classification. a**, The process of human visual perception. Cone cells are mainly responsible for color vision, while rod cells are primarily responsible for detecting shapes and movement in dim environments. The brain makes decisions based on the information transmitted from the retina. **b**, Biomimetic hyperspectral image classification process of HyPIS. Cos-encoder and sin-encoder convert spectral signals into a single point in the phasor plot, functioning similarly to cone cells in the retina and responsible for processing spectral information. SPI reconstructs the spatial information of objects, analogous to the role of cone cells in the retina. Cos-encoder: an encoder performs wavelength-dependent cosine-encoding; Sin-encoder: an encoder performs wavelength-dependent sine-encoding; SPD: single-pixel detector. **c**, The perceiving results of an object, derived from human visual recognition and cerebral processing. Two colors with similar spectra fail to be classified. **d**, Classification results of the same object when using HyPIS. Two colors with similar spectra can be recognized and classified.

Humans possess multisensory capabilities that enable them to perceive and understand complex and dynamic environments, with visual information accounting for a substantial portion of the brain's information-processing resources[47]. As illustrated in Fig. 1a, external visual scenes characterized by spatial, spectral, and temporal dimensions are captured by the eyes and conveyed to the retina for early-stage neural processing[48]. During this process, cone photoreceptors mediate high-acuity and trichromatic color vision through three spectrally distinct cone types, whereas rod photoreceptors primarily support highly sensitive luminance perception under low-light (scotopic) conditions and contribute indirectly to motion processing via magnocellular pathways[49]. Subsequently, higher-level visual and associative cortical areas integrate the retinal signals to support perception, decision-making, and the accumulation of visual experience. In this respect, the hyperspectral information-processing framework of HyPIS exhibits a functional resemblance to the hierarchical mechanisms of human visual perception. As illustrated in Fig. 1b, HyPIS illuminates an object with a sequence of encoding patterns. Two optical encoders perform wavelength-dependent sine (Sin-encoder) and cosine (Cos-encoder) encoding, recording three sets of intensity values: with Cos-encoder, with Sin-encoder, and without any encoding. Using these measurements and the known encoding patterns, the Phasor-SPI reconstruction algorithm computes the point coordinates (G, S) of the object in a phasor plot. In this scheme, the spectral information is converted into single phasor points, each corresponding to a different color, similar to the color perception function of cone cells. At the same time, SPI is used to recognize the shape and motion of objects, functioning similarly to the rod cells. By combining hyperspectral phase analysis with SPI, HyPIS achieves compressed sampling in both spatial and spectral dimensions, allowing classification results to be generated directly from the reconstructed G and S values, eliminating the need for full 3D hyperspectral data. It is worth noting that HyPIS outperforms human vision in classification tasks. Through comprehensive spectral coverage and precise spectral analysis, it can extract deeper environmental information. As shown in Figs. 1c and 1d, even when two spectrally similar colors cannot be distinguished by the human eye, HyPIS can achieve perfect recognition, enabling the identification of metameric colors.

**Results and discussion**

**Demonstration of HyPIS for classification with public datasets**

To demonstrate the concept of HyPIS, we first performed numerical verification with simulation data. We first conducted a comparison between the hyperspectral image classification process of HyPIS and the traditional hyperspectral image classification process. Figure 2a illustrates the hyperspectral image classification process of HyPIS. It can directly achieve hyperspectral image classification by compressed processing spatial-spectral 2D information based on phasor and SPI, which is a sensing and computing integrated device. The first line in Fig. S1 shows the processing flow of HyPIS. In HyPIS, a structured light field first enables spatially encoded illumination of the object, while an inexpensive single-pixel detector (SPD) collects the total intensity of the object after secondary encoding (i.e., spectral encoding of the total light intensity), as shown in Fig. S1a. Using the reconstruction algorithm proposed herein (see "Principles of HyPIS"), two images representing the G and S coordinate maps in the phasor plot are reconstructed, as depicted in the left panel of Fig. S1b. The right panel of Fig. S1b visualizes the storage usage for storing these two maps, where blue indicates unused memory and orange denotes used memory, clearly demonstrating the minimal storage requirement of HyPIS. Leveraging the reconstructed G and S images, the object's representation in the phasor plot is obtained, as shown in Fig. S1c. Classification is then directly achieved by partitioning regions based on coordinate positions in the phasor plot, without requiring additional classification algorithms. For comparison, Figure 2b illustrates the traditional hyperspectral image classification process. As described above, traditional hyperspectral image classification employs a "capture-storage-compute" distributed framework. It involves imaging using a hyperspectral camera, storing the data in memory, and using algorithms to perform post-processing. The second line in Fig. S1 shows the processing flow of the traditional method. A hyperspectral camera is used to probe the scene of interest, as shown in Fig. S1d. The acquired 3D hyperspectral data are then stored, depicting the collected hyperspectral dataset (left) and its memory footprint (right), as shown in Fig. S1e. Notably, the storage requirement is $N/2$ times that of HyPIS, where $N$ denotes the number of spectral bands. Classification is achieved using dedicated hyperspectral classification algorithms, as shown in Fig. S1f. By contrast, HyPIS offers three key advantages: 1) compressed sampling, 2) lower storage capacity, and 3) direct classification.

To quantitatively assess the performance of HyPIS, we utilized HyPIS to conduct simulations on public hyperspectral datasets, as shown in Fig. 2c-e. The CAVE dataset[50], which captured 31-band multispectral images at a resolution of 64×64 in the wavelength range of 400-700 nm at 10 nm intervals, is used as the hyperspectral data. In order to ensure consistency with the spectral range of the public datasets, we simulate a set of sine and cosine encoders operating in the 400-700 nm wavelength range, as shown in Fig. S2. The hyperspectral raw data are shown within the green dashed box in Figs. 2c-e. The reconstructed images of G and S (64×64 resolution) for these simulated samples, obtained using the HyPIS method under a 100% sampling rate, are shown within the pink dashed box in Figs. 2c-e. Within the orange dashed box are displayed the reconstructed phasor plots for each dataset. The simulated objects are represented as coordinate point sets in the phasor spectral plots, where point sets of the same color indicate that the objects have similar spectral curves. Based on this 2D phasor plot, direct classification of hyperspectral images can be achieved through geometric partitioning. The classification results for each object are shown within the blue dashed box in Figs. 2c-e, where the same color in the plot indicates objects of the same class, which is essentially determined by the spectral. It should be noted that HyPIS does not require acquiring 3D hyperspectral data of the scene under test during practical operation. The raw data within the green dashed box are used to simulate an object with spectral information. These simulation results demonstrate that the stored data volume is reduced to 1/15.5 of the raw data. This substantial data reduction efficiency not only minimizes storage and computational burdens but also facilitates real-time processing in resource-constrained environments, such as portable imaging devices or remote sensing systems.

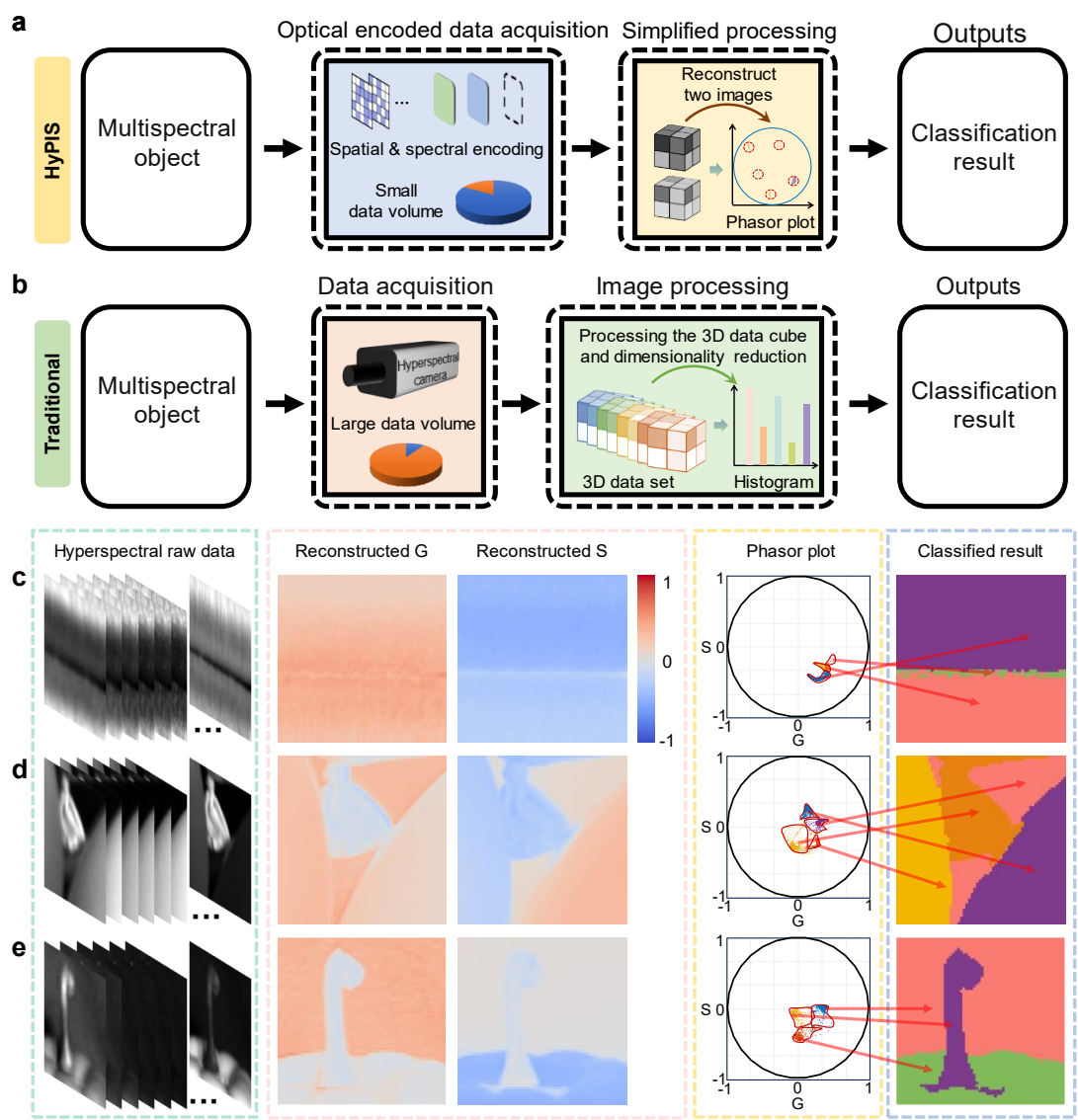

**Fig. 2. Comparison between the HyPIS and the traditional method hyperspectral image classification process. a**, Schematic illustration of the hyperspectral image classification process of HyPIS, which realizes the compressed processing of spatial-spectral 2D information, and directly achieves hyperspectral image classification. **b**, Traditional hyperspectral image classification process involves imaging using hyperspectral camera, storing a large 3D dataset, and performing post-processing with algorithms. Simulation classification results of HyPIS on public datasets: **c**, Thread spools; **d**, Balloons; **e**, Fake and real food.

**Experimental setup of HyPIS**

The experimental setup of our system is shown in Fig. 3a. A white LED is used as the light source. A high-speed digital micro-mirror device (DMD) generates a series of orthogonally discrete cosine modulation patterns that would project on the object plane. The light from the object, after passing the collecting lens, is detected by three single-pixel detectors (SPDs), which enables the simultaneous detection of three light beams. The first SPD serves as the reference detector and detects the light intensities without any optical encoding, while the other two detect them after the spectral modulation of the sine encoding (Sin-encoding) and cosine encoding (Cos-encoding), respectively. By correlating the modulation patterns and the three groups of measured single-pixel intensities, the 2D phasor plot of the object can be reconstructed. Conventional methods typically acquire spectral data using detector arrays, such as line-array detectors commonly employed in commercial spectrometers, as shown in Fig. 3b. In comparison, HyPIS encodes the spectra into the phasor domain with two custom-built encoders. The transmission spectra of the ideal encoders are shown in Fig. 3(c) and (d), which exhibit cosine and sine profiles in the 400-600 nm range. Then, each spectrum is spectrally encoded with such optical encoders during the measurement process. After the detection of SPDs, the phasor coordinates of these spectra can be directly acquired, as shown in Fig. 3e. All the measured spectra would be represented as unique point locations within the circular range, whose boundary corresponds to the narrowband spectra in the range of 400-600 nm. More simulation results can be found in Fig. S3. Therefore, the application of such spectral phasor encoding enables the classification of different spectra, which correspond to different regions of objects, without acquiring the entire spectral dataset, but only three single-point measurements.

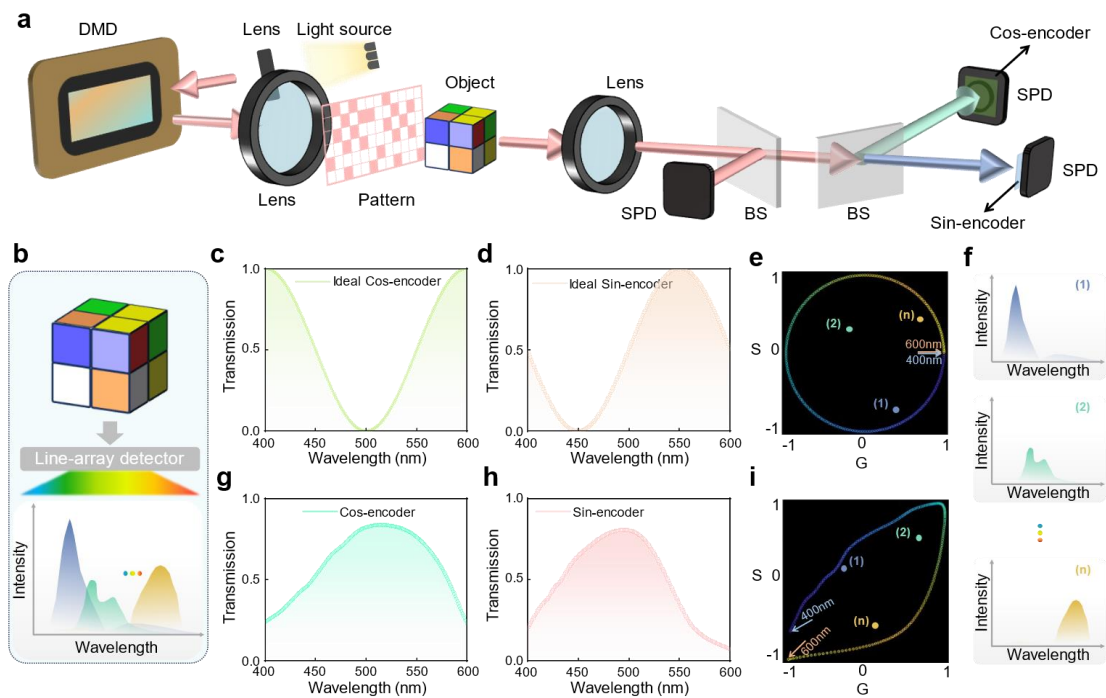

**Fig. 3. Experimental setup of HyPIS and characterization of the optical encoders. a**, Schematic of the experimental setup. DMD: digital micromirror device; BS: beam splitter; SPD: single-pixel detector. **b**, Illustration of traditional methods that measure the spectra of an object with a spectrometer with a line-array detector. **c-d**, Transmission spectra of the ideal phasor encoders, with the shape of cosine and sine, respectively. **e**, Phasor representation of the ideal encoders and different spectra of the object (**f**) are coded into unique positions within the phasor domain. **g-h,** Transmission spectra of the cosine and sine encoders used in this work. **i**, Phasor representation of the used encoders and the spectra from the object.

For convenience, here we adopt two generalized phasor encoders in the experiments, whose transmission spectra are presented in Figs. 3(g) and (h) were calibrated with a commercial spectrometer with 0.2 nm resolution. Though these spectra deviate from the ideal cosine or sine profiles, the phasor analysis and presentation can still be performed, as shown in Fig. 3i. The unknown spectra can also be represented within the ellipse-like phasor boundary. It should be noted that the 2D phasor plot can be further corrected into the circular representation using a correction factor (Fig. S4). However, the uncorrected elliptical phasor plot still exhibits clear representations, which are directly used in the following experiments without further correction.

## Experimental results

### HyPIS for Hyperspectral Image Classification

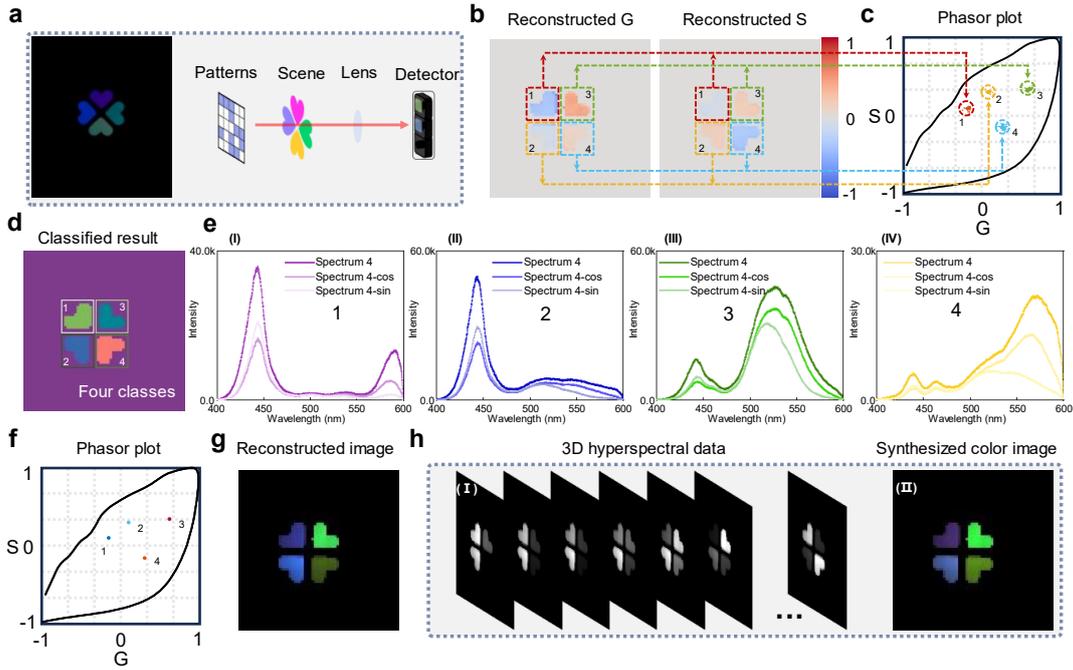

**Fig. 4. Demonstration of HyPIS with a transmission object. a**, Photograph of the used object, which is a four-leaf clover-like transmission mask. **b**, Reconstructed images of G and S components, respectively, where the two intensity values of each spatial location serve as the coordinates of its phasor point in **c**. **d**, Classification result according to phasor locations. **e**, Spectra of four areas measured using a spectrometer under three conditions: with Sin-encoder, with Cos-encoder, and without any encoder. **f**, 2D phasor plot of the measured spectra in **e**. **g**, Reconstructed pseudo-color image using HyPIS. **h**, (I): The 3D hyperspectral data that was constructed by combining the spectral data measured by a spectrometer with the images reconstructed by traditional SPI; (II): synthesized pseudo-color image.

To demonstrate HyPIS, we first performed the experiments with a four-leaf clover-like transmission object. Figure 4a shows the photo image of the used object that was stuck with different color filters, and also a simplified experimental procedure shown in the right part. The reconstructed images corresponding to the G and S dimensions in the phasor plot are presented in Fig. 4b, in which the intensities range from -1 to 1. Then the intensity values of each spatial pixel define the coordinates of its locations in the phasor domain in Fig. 4c. The average coordinate values corresponding to the four parts are (-0.16±0.02, 0.09±0.02), (0.09±0.02, 0.31±0.01), (0.60±0.03, 0.34±0.02), and (0.28±0.02, -0.15±0.03), respectively. The different spatial locations with similar spectral characteristics, shown in the square regions in Fig. 4b, are clustered within neighboring areas denoted by the dashed circles in Fig. 4c. This clustering behavior demonstrates the effective spectral discrimination of different regions within the sample. Based on this, we can directly obtain the classification results of the hyperspectral image of the target from the two graphs, G and S. Figure 4d shows the classification result, in which the four regions are divided into four classes, each of which has the same transmission spectra. It is worth noting that the classification results obtained through HyPIS are fully consistent with those of the traditional approach, which first acquires complete spectral information and then applies algorithms for classification. In order to verify this, we employed a spectrometer to measure the spectra of four distinct heart-shaped regions on the sample. Spectral data were acquired for each location under three conditions: with Sin-encoder, Cos-encoder, and without any encoder. The measurement results are presented in Fig. 4e, where subpanels (I)-(IV) correspond to the spectral profiles of the upper-left (Number 1), lower-left (Number 2), upper-right (Number 3), and lower-right (Number 4) regions of target, respectively. Leveraging the phasor analysis algorithm, we calculated the coordinates of the phasor points associated with these four sets of spectral curves. The computed results are shown in Fig. 4f, with the phasor point coordinates for the upper-left (Number 1), lower-left (Number 2), upper-right (Number 3), and lower-right (Number 4) heart-shaped regions of the four-leaf clover being (−0.15, 0.09), (0.10, 0.30), (0.61, 0.34), and (0.31, −0.16), respectively. It can be seen that the results obtained from HyPIS are basically consistent with those of spectral phasor analysis, which uses the data measured by the spectrometer. These results highlight the efficiency and simplicity of the proposed HyPIS method.

HyPIS achieves hyperspectral image classification through compressive sampling in spectral-spatial domain. It uses phasor encoding to map 1D spectral data to different points in the phasor plot, with each point corresponding to a specific color defined by the color space, and then employs SPI to process and reconstruct the spatial information of the object. This unique approach not only enables HyPIS to perform hyperspectral image classification directly but also to reconstruct high-quality full-color images of the target object. Figure 4(g) shows the reconstructed color image. For comparison, we reconstructed the hyperspectral image using a simplified approach by combining the spectra data with the images obtained by single-pixel imaging, as shown in Fig. 4h(I). Then, we synthesized pseudo-color using the CIE standard color synthesis method (Fig. 4h(II)), and it can be seen that this color is very close to the reconstructed image in Fig. 4(g). Notably, the data volume used by conventional approaches for color image reconstruction or image classification is 500 times that of the HyPIS scheme, consuming much more memory resources. More comparison of memory usage can be found in Fig. S5. Moreover, we validated the performance of HyPIS under different light distributions and detector gains, with the experimental results shown in Fig. S6. The results indicate that the hyperspectral phasor imaging and classification outcomes of HyPIS are not affected by the light distribution of the source or the detector gain level. Further simulations and experiments with the colorimetric card can be found in

Figs. S7 and S8 in the Supporting Information. Thus, our experimental and simulation results indicate that HyPIS achieves hyperspectral image classification and high-fidelity full-color images by performing compressed sampling and reconstruction simultaneously in both spatial and spectral dimensions, reducing data storage by two orders of magnitude.

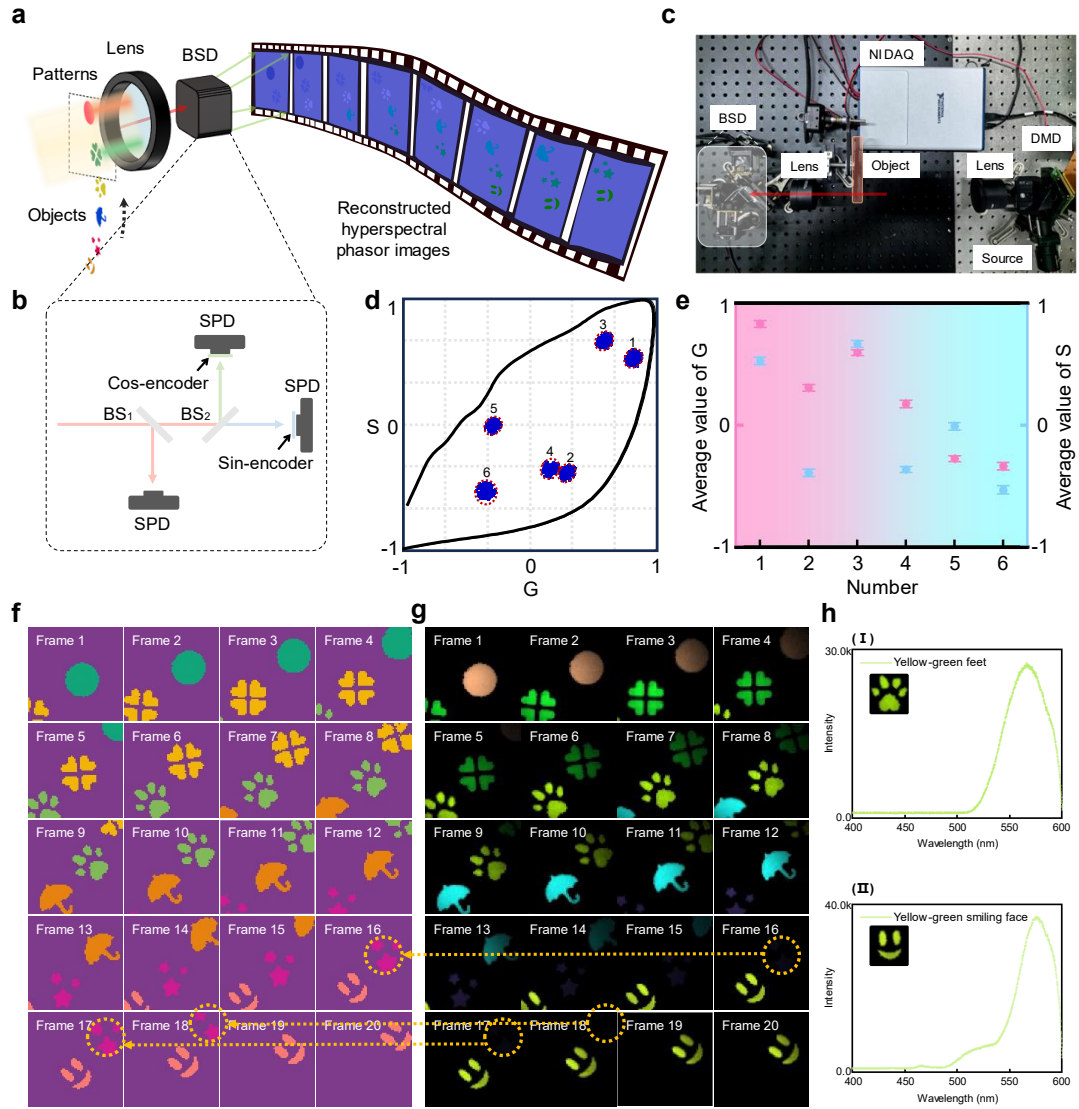

**Fig. 5. Demonstration of HyPIS for real-time hyperspectral phasor imaging and classification. a**, Illustration of HyPIS for real-time imaging. The encoded patterns were projected onto the moving object, where a black arrow indicates the direction of the object's motion. A beam-splitting detection (BSD) system is adopted to acquire the spectral encoded light intensities. **b**, Schematic diagram of the BSD. **c**, Photograph of the experimental setup, in which the object is mounted on a rotating stage. **d**, 2D phasor plot of the six different objects that have different transmission spectra. **e**, Statistics of *G* and *S* values in the phasor domain for the six different objects, along with their standard deviations. **f**. Classification results of the objects and the background, showing in different colors. **g**, Reconstructed pseudo-colored images using HyPIS. **h**, Transmission spectra of the two example objects that have slight differences in the spectral profiles.

To demonstrate the classification performance of HyPIS in dynamic motion scenarios, we further conducted the experiments for real-time imaging. As illustrated in Fig. 5a, the object used in this experiment is a dynamic rotating mask consisting of six transmission objects, each with a distinct shape and spectral transmittance. A compact beam-splitting detection system is applied for the spectral phasor encoded detection, as presented in Figs. 5b and 5c. Using HyPIS, we reconstructed 160 frames of the dynamic scene with a size of 64 × 64 pixels and the frame rate of approximately 2.68 fps. Figures. S9 displays 20 frames of reconstructed images from the G and S videos. The acquired 2D phasor plots of the six objects are shown in Fig. 5d, in which identical objects across each frame are consistently mapped. Figure 5e presents the averaged G and S values of the phasor coordinates for each object, along with their standard deviations. We then performed the image classification, based on the spectral features shown on the phasor plot. Figure 5f shows 20 example frames with an interval of 2.98 seconds between each frame, demonstrating that the objects within the scene are categorized into six classes. Figure 5g further presents the reconstructed pseudo-color images. The video synthesized from these 20 frames is presented in Video S1 of the Supporting Information.

A gradient of decreasing illumination intensity from the lower-left to lower-right corner is evident in the red circle across Frames 1-4 in Fig. 5g. The intensity of the red circle in Frame 4 decreased due to low illumination intensity. However, such non-uniform illumination had no discernible impact on classification results in Fig. 5f. Similarly, the purple star in Frames 16-18 of Fig. 5g, marked by orange dashed circles, becomes visually imperceptible, nearly imperceptible. It is still classified as a distinct category in the corresponding results of Fig. 5f, demonstrating that HyPIS can robustly segment and classify objects even under extremely low visibility conditions. Moreover, the feet in frames 4-12 and the smiling face in frames 13-20 of Fig. 5g exhibit the same yellow-green chromaticity, which makes direct classification by these color images unfeasible. However, the 2D phasor plot (Fig. 5d) clearly supports their classification, as confirmed by the classification results in Fig. 5f, where the feet and smiling face are labeled in green and pink, respectively. The corresponding transmission spectra (Fig. 5h) reveal subtle spectral differences that lead to distinct phasor coordinates (labels 2 and 4 in Fig. 5d). This highlights HyPIS's effectiveness in hyperspectral imaging–based classification and its capacity to differentiate metameric objects with similar apparent colors, underscoring a fundamental advantage of hyperspectral imaging techniques.

**HyPIS for Scene Recognition**

Another advantage of HyPIS is its potential for compact and lightweight hardware design, enabling easy integration into mobile platforms such as UAVs for real-time hyperspectral classification and object recognition (Fig. 6(a)). HyPIS can be further extended to not only achieve classification based on hyperspectral phasor imaging but also the recognition tasks based on the pre-calibrated phasor database. HyPIS can be extended to perform both hyperspectral phasor imaging–based classification and object recognition using a pre-calibrated phasor database (See Fig. S10 in the Supplementary Information for more details about the classification and recognition process), as shown in Fig. 6. Figures 6b and 6c show the experimental setup and the used object (an

Anthurium leaf).

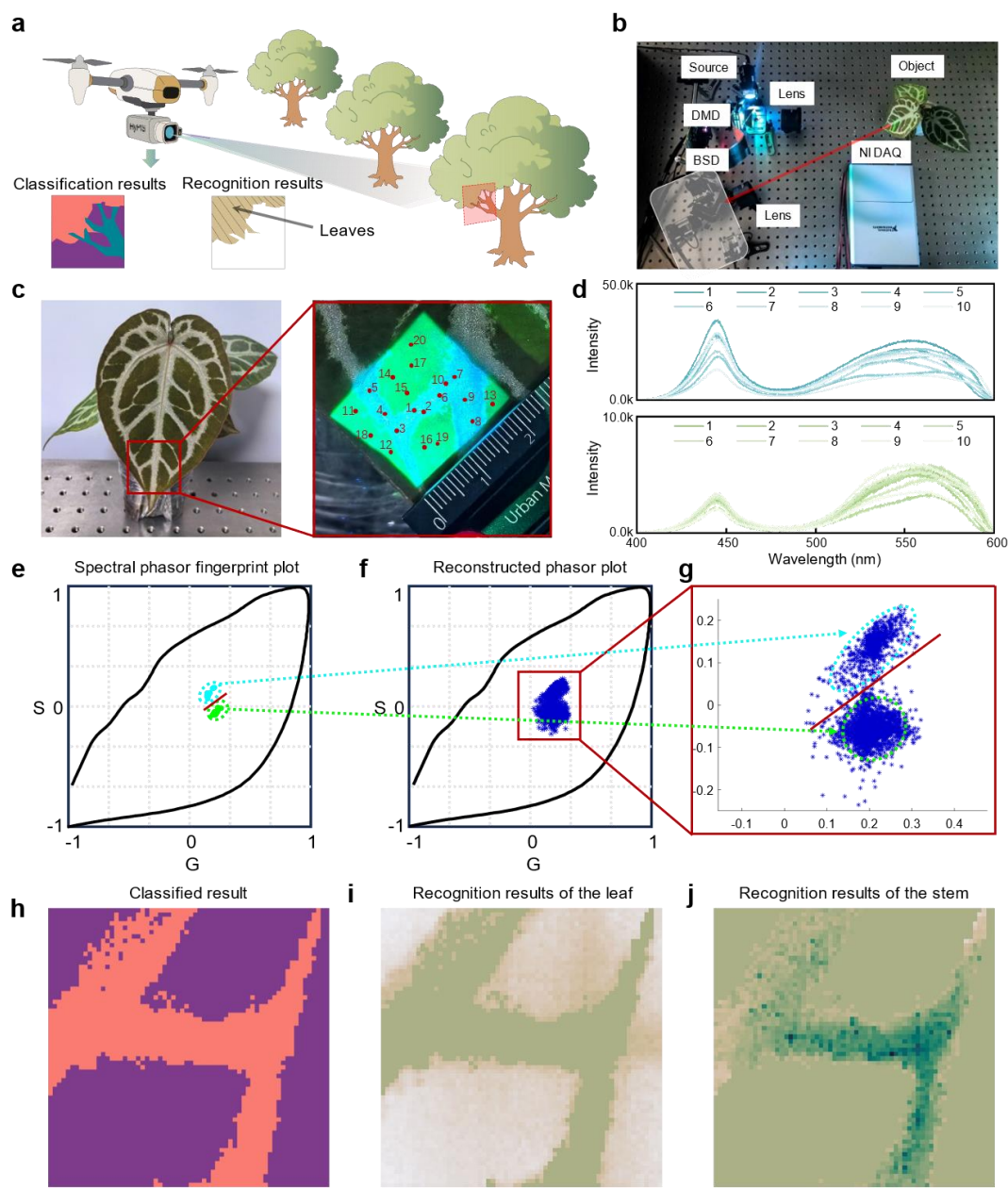

**Fig. 6. Demonstration of HyPIS with a leaf as the object. a**, Illustration of HyPIS deployed on a small unmanned aerial vehicle. **b**, Photograph of the HyPIS experimental setup in the reflective configuration. **c**, Photograph of the object—an Anthurium leaf. The right image shows the illuminated detection area (~2.2 cm × 2.2 cm). **d**, Measured reference spectra at selected points labeled in (c). **e**, Referenced spectral phasor plot of the object, derived from data in (d). **f**, 2D phasor plot retrieved by HyPIS. **g**, Enlarged view of the red dashed region in (f). **h**, Classification result obtained using HyPIS. **i** & **j**, Segmented regions corresponding to the leaf and stem areas indicated in (c).

We first measured the spectra corresponding to the positions of 20 points, as indicated in Fig. 6c, using a commercial spectrometer. The corresponding results are shown in Fig. 6d, with labels 1–10 representing points on the stem and labels 11–20 representing points on the leaves. Using the traditional phasor algorithm, the spectral phasor fingerprints of the Anthurium were calculated (Fig. 6e), revealing two distinct clusters corresponding to stems and leaves due to their spectral differences. Figure 6f shows the 2D phasor plot obtained by HyPIS, where each point represents a point location of the detected area. Figure 6g provides an enlarged view of the red dashed region in Fig. 6f, where the cyan and green dashed denotes the stem and leaf regions, based on the pre-calibrated data in Fig. 6e. Based on this, the positions of the stem part and the leaf part can be classified and recognized. The classification results are shown in Fig. 6h, and Figs. 6i and 6j present the recognized leaf and stem areas, respectively. These results closely match the corresponding regions in Fig. 6c, demonstrating that HyPIS can effectively perform classification and recognition based on hyperspectral phasor imaging.

**Conclusion**

In this work, we present HyPIS, a hyperspectral phasor imaging system using single-pixel detectors that enables direct spectral task-oriented without acquiring the full hyperspectral datacube. It achieves compression sampling in the spatial dimension through structured illumination and in the spectral dimension using wavelength-dependent sine and cosine encoding, enabling simultaneous compression in both spatial and spectral dimensions, reducing the amount of data obtained by up to two orders of magnitude. Owing to this dual-domain compression, HyPIS significantly reduces data throughput and system complexity while preserving discriminative information required for spectral task. We verified the effectiveness of the proposed method with both transmission and reflective objects and demonstrated the dynamic imaging of HyPIS to classify six different objects, including metameric targets, at 2.68 fps. In addition, HyPIS is extended to perform object recognition based on a pre-calibrated spectral phasor database, as demonstrated using a plant target, highlighting its capability for data-efficient processing without spectral acquisition.

Our work demonstrates that HyPIS is a task-oriented sensing strategy, which aligns with recent advances in intelligent spectral machine vision, where spectral inference is performed directly within the sensing hardware without generating full hyperspectral 3D data cubes[19]. HyPIS realizes this paradigm through optical encoding, and further extends it to a single-pixel architecture. In the future, researchers can achieve HyPIS-like spectral task devices by designing metasurfaces[51], low-dimensional materials[52], and photodetectors[53, 54] with Phasor encoding or optical responses. Moreover, it can be readily integrated with deep neural networks[55, 56] to support high-resolution inference under extreme measurement constraints, and can be extended to invisible spectral bands such as the near-infrared and terahertz regimes. We anticipate that HyPIS will open new opportunities for the development of portable, low-cost, and energy-efficient spectral task sensing platforms, particularly for applications where hardware simplicity and real-time decision-making are critical.

**Methods**
**Principles of HyPIS**

In SPI, an object $O(x_0, y_0)$ is illuminated sequentially by $N$ encoding patterns $P_i(x_0, y_0)$, the total intensities $y_i$ of the corresponding total reflected or transmitted light, measured by an SPD, can be expressed as:

$$y_i = \Sigma_{x_0,y_0} P_i(x_0, y_0) \cdot O(x_0, y_0), \quad (1)$$

where $i = 1, 2, 3 \cdots, N$ is the index of the pattern, and $x_0$ and $y_0$ are spatial coordinates. It should be noted that $y_i$ contains the spectral response of the object to the entire light source spectrum. When filtering the light intensity through an ideal narrowband filter (with a central wavelength of $\lambda_1$), the total intensities $y_i$ equals the total light intensity $y_{\lambda_1}$ of the object secondarily encoded at $\lambda_1$. Therefore, $y_i$ can be expressed as:

$$y_i = y_{\lambda_1}. \quad (2)$$

When the narrowband filter has two central wavelengths of $\lambda_1$ and $\lambda_2$, the total intensities $y_i$ equals the sum of the light intensities of the object secondarily encoded at each of these two wavelengths. It can be expressed as:

$$y_i = y_{\lambda_1} + y_{\lambda_2}. \quad (3)$$

When no filters are used, the total intensity $y_i$ can be also expressed as:

$$y_i = \Sigma_\lambda y_{\lambda i}(\lambda), \quad (4)$$

where $\lambda$ denotes the wavelength range covered by the light source. Eq. 4 denotes that the intensity value equals the sum of intensities of the object at different wavelengths encoded by the patterns. This theory provides theoretical support for implementing a hyperspectral imaging scheme based on a single-pixel detector through spatial-spectral dual-dimensional encoding.

Different from traditional SPI, HyPIS not only reconstructs the spatial information of objects but also directly enables hyperspectral image classification. In HyPIS, three sets of intensity values are recorded: with the Cos-encoder, with the Sin-encoder, and without encoder. According to Eq. 4, when the Cos-encoder is placed in front of the single-pixel detector to work as a spectral encoder, the total intensity $y_{icos}$ can be expressed as:

$$y_{icos} = \Sigma_\lambda y_{i\lambda}(\lambda) \cdot I_{cos}(\lambda), \quad (5)$$

where $I_{cos}(\lambda)$ represents the transmittance curve of cos-encoder. To retrieve the image $O(x_0, y_0)_{cos}$ after passing through the cos-encoder, a compressive sensing algorithm is used. Assume the image $O(x_0, y_0)_{cos}$ has a sparse representation under a 2D discrete cosine transform (2D-DCT) basis:

$$O(x_0, y_0)_{cos} = \Sigma_{u,v} s_{u,v} \cdot \psi_{u,v}(x_0, y_0), \quad (6)$$

where $\psi_{u,v}(x_0, y_0)$ is the 2D-DCT basis function and $s_{u,v}$ are the sparse coefficients. Then, the image is reconstructed by solving the following optimization problem:

$$argmin_s \frac{1}{2} \Sigma_{i=1}^N \left(y_{icos} - \Sigma_{x_0,y_0} P_i(x_0, y_0) \cdot \Sigma_{u,v} s_{u,v} \psi_{u,v}(x_0, y_0)\right)^2 + \lambda \Sigma_{u,v} |s_{u,v}|, \quad (7)$$

where $\lambda > 0$ is is the regularization parameter controlling the sparsity strength, and $|\cdot|$ denotes the absolute value. After solving this optimization problem to obtain the optimal sparse coefficients $s_{u,v}$, the image $O(x_0, y_0)_{cos}$ can be obtained according to Eq. 6.

There is an analogous reconstruction process for retrieving the image encoded by the Sin-encoder $O(x_0, y_0)_{sin}$ and the unencoded image $O(x_0, y_0)$. Then, with the help of phasor analysis, the point coordinates $G(x_0, y_0)$ and $S(x_0, y_0)$ of the object $O(x_0, y_0)$

in the phasor plot can be obtained with:

$$G(x_0, y_0) = 2 * \frac{\frac{O(x_0,y_0)_{cos}}{O(x_0,y_0)} - I_{cos-min}}{I_{cos-max} - I_{cos-min}} - 1, \qquad (8)$$

$$S(x_0, y_0) = 2 * \frac{\frac{O(x_0,y_0)_{sin}}{O(x_0,y_0)} - I_{sin-min}}{I_{sin-max} - I_{sin-min}} - 1, \qquad (9)$$

where $I_{cos-max}$ and $I_{cos-min}$ are the the maximum and minimum transmittance values of the Cos-encoder respectively, and $I_{sin-max}$ and $I_{sin-min}$ are the the maximum and minimum transmittance values of the Sin-encoding respectively. Finally, using this theory, we can obtain the whole phasor plot of the object.

**Experimental apparatus**

A white LED (GCI-060411, Daheng Optics) is used as the light source, with the illumination light in the range of 400-600 nm that is filtered by a bandpass (MEFH10-600SP, LBTEK). The light source is first shaped into collimated parallel light through a lens (GCL-M010158N, Daheng Optics) and illuminates a high-speed DMD (GmbH V-7001, ViALUX), generating a series of orthogonally grayscale modulation patterns with a refresh rate of 22.0 kHz. We use the spatial dithering method to generate the patterns via DMD, as shown in Fig. S11. The patterns feature a $64 \times 64$ superpixel resolution. Each superpixel contains $12 \times 12$ micromirrors, spanning the central $768 \times 768$ pixels of the DMD. A lens L1 (MAD507-AB, LBTEK) is used to project the encoding patterns onto the object plane. Then, the light focused by a lens L2 (MAD508-AB, LBTEK), resulting from the interaction with the object, including transmission, reflection, and scattering, is successively split using a beam splitting detection (BSD) system. This system consists of two beam splitters (BS1 (BS2137-A, LBTEK) and BS2 (BS2155-A, LBTEK)) and three SPDs (PDAPC2, Thorlabs), which enable the simultaneous detection of the intensities of three light beams. Three SPDs are placed in three optical paths, respectively, to collect the light intensity. A data acquisition module (USB-6353, National Instruments) synchronizes the DMD and SPD to achieve data acquisition.

The experimental samples used in Fig. 5 and Fig. 6 were fabricated by stamping a pattern onto black paper using a press stamp. For Fig. 5, the patterned regions were hollowed out to permit light transmission, forming a four-leaf clover shape, whereas the non-patterned areas blocked light passage. Subsequently, four distinct color filters were individually affixed behind the hollowed sections of the four-leaf clover. For Fig. 6, we used six differently shaped stamps to imprint patterns onto a circular black turntable. Then, we attached six distinct colored filters behind the hollowed patterns of the stamps to create the experimental sample. In the experiment, the turntable was driven by a compact rotary mount (DDR25/M, Thorlabs).

**Acknowledgements**

This work was supported by the National Natural Science Foundation of China (62503032, U23A20481, 62275010, 62573029), and the China Postdoctoral Science Foundation (2025M780804). We are grateful to the Atomic-Scale In Situ Fabrication